\documentclass{ws-mpla}

\newcommand{\vct}[1]{\bm{#1}}
\newcommand{\ket}[1]{|#1\rangle}

\newcommand{\mtrix}[3]{\langle \,#1\,|\,#2\,|\,#3\,\rangle}
\newcommand{\kf}{k_{\rm F}}
\newcommand{\ekf}{E_{\rm F}}
\newcommand{\kfp}{k_{\rm p}}
\newcommand{\kfn}{k_{\rm n}}

\newcommand{\vf}{v_{\rm F}}

\newcommand{\thp}[1]{\theta^{\rm (p)}_{\vct{#1}}}
\newcommand{\thn}[1]{\theta^{\rm (n)}_{\vct{#1}}}

\newcommand{\nnbar}{{\rm N}\overline{\rm N}}
\newcommand{\nbar}{\overline{\rm N}}

\begin{document}

\markboth{Kurasawa, Suzuki}
{Gamow-Teller Sum Rule}

\catchline{}{}{}{}{}

\title{
GAMOW-TELLER SUM RULE\\[0.3em] IN RELATIVISTIC NUCLEAR MODELS
}

\author{\footnotesize HARUKI KURASAWA}
\address{Department of Physics, Faculty of Science, Chiba University, Chiba
263-8522, Japan}

\author{\footnotesize TOSHIO SUZUKI}
\address{Department of Applied Physics, Fukui University,
Fukui 910-8507, Japan}

\maketitle


\begin{abstract}
Relativistic corrections are investigated
to the Gamow-Teller(GT) sum rule with respect to the difference
between the $\beta_-$ and $\beta_+$ transition strengths in nuclei.
Since the sum rule requires the complete set of the nuclear states,
the relativistic corrections come from the anti-nucleon degrees of
freedom. In the relativistic mean field approximation,
the total GT strengths carried by the nucleon sector
is quenched by about 12\% in nuclear matter, while by about 8\%
in finite nuclei, compared to the sum rule value. The coupling
between the particle-hole states with the nucleon-antinucleon states is
also discussed with the relativistic random phase approximation,
where the divergence of the response function is renormalized with use
of the counter terms in the Lagrangian.
It is shown that the approximation to neglect the divergence,
like the no-sea approximation extensively used so far,
is unphysical, from the sum-rule point of view.

\end{abstract}


\section{Introduction}\label{i}
The sum rules play an important role in a wide range of physics.
Since they are derived from the fundamental principles of the theory,
their violation leads us to the reconstruction of the theory.
Recently, precise analyses of experiment have been performed
on the sum rule with respect to the difference
between the $\beta_-$ and $\beta_+$ Gamow-Teller(GT) transition
strengths in $^{90}$Zr.
It has been shown that the GT sum rule value is
quenched by 12\%$\pm$6\%\cite{yako}, not including an overall
normalization uncertainty in the GT unit cross section of 16\%.
The analyses have been performed
in the excitation-energy region up to 50 MeV.
Experimentally, it is not clear yet that the GT sum rule is violated, 
mainly because of the uncertainty in the GT unit cross section.
It is important, however, for nuclear physics to study theoretically
whether or not the GT sum rule is violated,
since it should be hold, if the nucleus is a non-relativistic
quantum mechanical system composed of nucleons.
The study of the violation is also important
for other several areas of physics.
For example, information of nuclei is necessary for 
neutrino physics or supernova physics, where the dominant processes are
governed by the GT and Fermi transitions.
The pion condensation in neutron star can not be also discussed
without the knowledge on the violation of the GT sum rule\cite{sts}. 
Correct understanding of nuclei makes it possible to discuss
the GT strength in different situations. 

Theoretically, 
it is pointed out that the GT sum rule is violated
owing to the non-nucleonic degrees of freedom in nuclei, that is
the $\Delta$ degrees of freedom.
The $\Delta$ is an excited state of the nucleon,
whose mass is 1323 MeV. The $\Delta$-hole states are excited with
additional GT strengths to the sum rule value. 
While the sum rule value is proportional to the difference
between the neutron and the proton number,
the total GT strength of
the $\Delta$-hole states is proportional to the number of
nucleons. Therefore, if there is a repulsive coupling of the nucleon
particle-hole states with the $\Delta$-hole ones,
their small mixing yields a considerable reduction of the GT strength
in a low excitation-energy region\cite{ss}. 
Indeed, the coupling force is predicted to be repulsive.
Recent calculations have shown that
19(14)\% of the sum rule value is
taken by the $\Delta$-hole states, using the coupling constant of
Chew-Low (the constituent quark) model\cite{bentz}.

There are more naive corrections to the GT sum rule of the nucleonic
degrees of freedom in the non-relativistic theory. 
In nuclei, relativistic corrections have been believed to be about 10\% 
for a long time, since the nuclear Fermi velocity is estimated to be
about 0.27 from the nuclear density in non-relativistic models.
In the case of the sum rule, this correction is
related to  a requirement of the complete set of nuclear states.
In a relativistic framework,
the complete set is composed of the nucleon(${\rm N}$)
and the anti-nucleon($\nbar$) states.
Since $\nnbar$ states are at the time-like region above 1
GeV, the GT strengths in the lower energy region
should be reduced. The sum rule value may not be changed, but a part of the
GT strengths are taken by $\nnbar$ states, which can not be excited
through usual charge-exchange reactions. This means the violation of
the GT sum rule in the nucleonic degrees of freedom\cite{ksg1}. 

The purpose of the present paper is to review recent work on the
relativistic corrections to the GT sum rule[5-11]. 
It will be shown that naive relativistic corrections reduce the GT
strengths in the low excitation-energy region
by about 12\% in nuclear matter with the normal density
and about 8\% in finite nuclei.
We will also examine effects of the coupling between the particle-hole
states and the nucleon-antinucleon states in the random
phase approximation by assuming relativistic Landau-Migdal parameters.
The coupling requires the renormalization of the
divergence in the relativistic response function. A model to renormalize
the divergence is presented.
It will be also shown that the so-called no-sea approximation\cite{fruns},
which is employed frequently in the literature\cite{ksg1,ks2,ma,maring},
is unphysical, from the sum-rule point of view.
 
\section{Gamow-Teller Sum Rule}
Let us briefly mention the GT sum rules in the non-relativistic and
relativistic theory. They are the model-independent sum rules
with regard to the difference between the total $\beta_-$ and $\beta_+$
transition strengths. The sum rule values are the same in both cases,
and given by the difference between
the  neutron number $N$ and the proton number $Z$ of the system.

\subsection{Gamow-Teller sum rule in the non-relativistic theory}
In the non-relativistic theory, the GT sum rule is obtained by
calculating\cite{osterfeld} 
\begin{eqnarray}
S_{{\rm nonrel}}=\sum_n\{\mtrix{0}{Q_+}{n}\mtrix{n}{Q_-}{0}- \mtrix{0}{Q_-}{n}
\mtrix{n}{Q_+}{0}\},\label{ngtsum}
\end{eqnarray}
where $Q_\pm$ denotes the $\beta_\pm$ transition operators, respectively,
\begin{eqnarray}
 Q_\pm = \sum_i^A\sigma_{yi}\tau_{\pm i}\, \quad 
\tau_\pm =\left(\tau_x \pm i\tau_y\right)/\sqrt{2}.
\end{eqnarray}
With use of the
closure property of the intermediate states,
the sum is expressed in terms of the commutator, so that the sum value
is calculated to be
\begin{eqnarray}
S_{{\rm nonrel}}=\mtrix{0}{[Q_+,\,Q_-]}{0}=2(N-Z).\label{ngtsumv}
\end{eqnarray}
Thus, the sum rule is just the result of the commutation relation
between the isospin operators. Hence,
any non-relativistic model should satisfy it.\footnote{It
should be noted that the sum rule value
is given as $3(N-Z)$ in the literature\cite{osterfeld},
where the excitation operator is defined by
$\sum \vct{\sigma}(\tau_x\pm i\tau_y)/2$, instead of $Q_\pm$.}

If there is no ground state correlations in $N>Z$ closed subshell
 nuclei, as sometimes assumed for $^{90}$Zr and $^{208}$Pb, we have
\begin{eqnarray}
Q_+\ket{0}=0.
\end{eqnarray}
Then the sum rule value is exhausted by the proton particle and neutron
hole states excited through the $\beta_-$ transitions,
\begin{eqnarray}
\sum_n\mtrix{0}{Q_+}{n}\mtrix{n}{Q_-}{0}=2(N-Z).\label{gtsum2}
\end{eqnarray}

\subsection{Gamow-Teller sum rule in relativistic theory}
In terms of the nuclear field $\psi(\vct{x})$,
the excitation operators of the $\beta_\pm$ transitions are written as  
\begin{eqnarray}
F_\pm=\int d^3x \, \overline{\psi}({\vct{x})}\varGamma_\pm\psi(\vct{x}),\label{f}
\end{eqnarray}
where $\Gamma_\pm$ stands for
\begin{eqnarray}
\varGamma_\pm=\gamma_5\gamma_y\tau_\pm.
\end{eqnarray}
The GT sum is described as
\begin{eqnarray}
S_{{\rm rel}}=\sum_n\left\{\mtrix{0}{F_+}{n}\mtrix{n}{F_-}{0}-
\mtrix{0}{F_-}{n}\mtrix{n}{F_+}{0}\right\}.\label{rgtsum}
\end{eqnarray}
With use of the closure property of the intermediate states, the above
equation is expressed by the commutator, as in the non-relativistic theory, 
\begin{eqnarray}
S_{{\rm rel}} =\mtrix{0}{[F_+,F_-]}{0}.\label{c}
\end{eqnarray}
The commutator is calculated as
\begin{eqnarray}
[F_+, F_-]&=&\int d^3xd^3x'(\gamma_0\Gamma_+)_{\alpha\alpha'}
 (\gamma_0\Gamma_-)_{\beta\beta'}\left[\psi^\dagger_\alpha(\vct{x})
\psi_{\alpha'}(\vct{x}), \psi^\dagger_\beta(\vct{x'})
\psi_{\beta'}(\vct{x'})\right]\\ \nonumber
&=&2\int d^3x\, \psi^\dagger(\vct{x})\tau_z\psi(\vct{x}),
\end{eqnarray}
by using the fact that
\begin{eqnarray}
\left\{\psi^\dagger_\alpha(\vct{x})
,\psi_{\alpha'}(\vct{x'})\right\}=\delta_{\alpha\alpha'}\delta(\vct{x}
-\vct{x'}).\label{ac}
\end{eqnarray}
Finally we obtain the GT sum rule in the relativistic theory as
\begin{eqnarray}
S_{{\rm rel}}=2(N-Z). \label{rgtsmv}
\end{eqnarray}
Thus, the sum rule value is the same as in the non-relativistic
theory. It should be noted, however, that Eq.(\ref{ac}) holds in
including the anti-nucleon degrees of freedom, and
the closure property used in Eq.(\ref{c}) requires $\nnbar$ states
in addition to particle-hole states.
If we neglect the $\nnbar$ states,
the sum value of the nucleon sector must be reduced by the relativistic
effect, compared to $2(N-Z)$. 

\section{Relativistic mean field approximation}
So far, relativistic corrections have been estimated with use
of the $p/M$ expansion of the free field.
Recent development of relativistic
nuclear models for the past 30 years, however,
make it possible to estimate the corrections
consistently with the nuclear wave function without the
expansion\cite{serot,ring}. 

 In the present section, we estimate relativistic corrections
 in the mean field  approximation.
 First, we study the GT transition in nuclear
matter, assuming that the nucleons and antinucleons are bounded in Lorentz
scalar $U_{\rm s}$ and vector $U_0$ potentials. Next, we will discuss
the relativistic corrections in finite nuclei, according to the more detailed
mean field approximation.

In nuclear matter in Lorentz scalar and vector potentials, the mean
field is given by the free field, but with the effective nucleon and
anti-nucleon mass $M^\ast=M-U_{\rm s}$, 
\begin{eqnarray}
 \psi(\vct{x})=\int\!\frac{d^3p}{(2\pi)^{3/2}}\sum_{\alpha}
\Bigl(
u_\alpha(\vct{p})\exp(i\vct{p}\cdot\vct{x})\,a_\alpha(\vct{p})
+v_\alpha(\vct{p})\exp(-i\vct{p}\cdot\vct{x})\,b^\dagger_\alpha(\vct{p})
\Bigr),
\end{eqnarray}
where we have defined
\begin{eqnarray}
u_\alpha(\vct{p})=u_\sigma(\vct{p})\,\ket{\tau}\,,\ \ 
( \alpha=\sigma, \tau )\,, \ \ {\rm etc}..\nonumber
\end{eqnarray}
The positive and negative spinors are given by
\begin{eqnarray}
u_\sigma(\vct{p})=\sqrt{\frac{E_p+M^{\ast}}{2E_p}}
\left(
\begin{array}{cc}
1\\
\dfrac{\vct{\sigma}\cdot\vct{p}}{E_p+M^{\ast}}
\end{array}
\right)\xi\,,\quad
v_\sigma(\vct{p})=
\sqrt{\frac{E_p+M^{\ast}}{2E_p}}
\left(
\begin{array}{cc}
\dfrac{\vct{\sigma}\cdot\vct{p}}{E_p+M^{\ast}}\\
1
\end{array}
\right)\xi\,. \nonumber
\end{eqnarray}
In the above equation, we have used abbreviations
$E_p=\sqrt{M^{\ast 2}+\vct{p}^2}$ and $\xi$ for Pauli spinor. 
The Lorentz vector potential does not appear explicitly in the field.
Using these notations, we have for $\beta_-$ and $\beta_+$
excitations in $N>Z$ nuclei, respectively,
\begin{eqnarray}
F_-\ket{0}&=&\sqrt{2}
 \int\!d^3p\,\sum_{\sigma\sigma'}
\Bigl(
\overline{u}_{\sigma}(\vct{p}) {\varGamma}u_{\sigma'}(\vct{p})\,
a^\dagger_{\sigma {\rm p}}(\vct{p})a_{\sigma'{\rm n}}(\vct{p})\Bigr.
\nonumber\\
& &
\phantom{\hspace{60pt}}
\left.
 +\,
\overline{u}_\sigma(\vct{p}) {\varGamma} v_{\sigma'}(-\vct{p})\,
a^\dagger_{\sigma{\rm p}}(\vct{p})
b^\dagger_{\sigma'{\rm n}}(-\vct{p}) \right)\ket{0},\nonumber \\
\noalign{\vskip4pt}
F_+\ket{0}&=&\sqrt{2}
\int\!d^3p\,\sum_{\sigma\sigma'}
\overline{u}_\sigma(\vct{p}){\varGamma} v_{\sigma'}(-\vct{p})\,
a^\dagger_{\sigma{\rm n}}(\vct{p})
b^\dagger_{\sigma'{\rm p}}(-\vct{p})\,\ket{0},\nonumber
\end{eqnarray}
$\Gamma$ being $\varGamma=\gamma_5\gamma_y$.
The above equations show that there are excitations of
proton-antineutron states in the $\beta_-$ transitions in addition to
particle-hole states, while neuron-antiproton states in the $\beta_+$
transitions. 
Using these equations, we obtain the total GT strengths for the
$\beta_- $ and $\beta_+$ transitions\cite{ksg1},
\begin{eqnarray}
\mtrix{0}{F_+F_-}{0} 
&=&
4\,\frac{V}{(2\pi)^3}\int\!\frac{d^3p}{E_p^2}\,
\left[
\thn{p}\left(1-\thp{p}\right)\left(M^{\ast2}+p_y^2\right) \right.
\nonumber \\
& &
\phantom{\hspace{60pt}} 
 \left.
+\,\left(1-\thp{p}\right)\left(\vct{p}^2-p_y^2 \right) \right],
\label{beta-}\\
\noalign{\vskip6pt}
\mtrix{0}{F_-F_+}{0} 
&=&
4\,\frac{V}{(2\pi)^3}\int\!\frac{d^3p}{E_p^2}\,
\left(1-\thn{p}\right)\left(\vct{p}^2-p_y^2 \right),\label{beta+}
\end{eqnarray}
where we have defined the step function and the volume of the system
as,
\begin{eqnarray}
\theta_{\vct{p}}^{(i)}=\theta(k_i -|\vct{p}|)\,\,\, {\rm for}\,\,
i=p \,\,{\rm and}\,\, n, \ \ \ 
V=A\left(3\pi^2/2\kf^3\right),\nonumber
\end{eqnarray}
$\kfn$ and $\kfp$ being Fermi momentum of neutrons and protons,
\begin{eqnarray}
k_n^3=\frac{2N}{A}\kf^3, \quad k_p^3=\frac{2Z}{A}\kf^3.
\end{eqnarray}
The analytic expressions in Eqs.(\ref{beta-}) and (\ref{beta+}) for
nuclear matter are useful for discussions of relativistic corrections
to the $\beta_-$ and $\beta_+$ transition strengths.
As seen from the step functions in the square brackets of
Eq.(\ref{beta-}), the first term stems from the transitions of neutrons
to proton states, while the second term is due to the transition
of anti-neutrons to proton states. In the second term, the step function
appears with a minus sign because of the Pauli principle.
The part with the step function is called frequently Pauli blocking term.
The integral of the first term is finite,
but the one of the second term is infinite, although Pauli blocking term
gives the finite value. 
For the $\beta_+$ transitions, Eq.(\ref{beta+}) yields the matrix
element coming from the transitions of anti-protons to neutron states only,
since we assume nuclei with $N>Z$. The integral of the right hand side
is also infinite.

The GT sum rule is obtained from the difference between
Eqs.(\ref{beta-}) and (\ref{beta+}).
Each of them is infinite,
but the infinite terms exactly cancel each other and
the difference becomes finite, yielding the GT
sum rule value, as in Eq.(\ref{rgtsmv})
\begin{eqnarray}
\mtrix{0}{F_+F_-}{0} - \mtrix{0}{F_-F_+}{0} = 2(N-Z).\label{rel}
\end{eqnarray}
The sum rule is thus satisfied, if we take into account the $\nnbar$
states.

The $\nnbar$ states can not be excited by usual charge-exchange
reactions, as mentioned before.
If there is no coupling between particle-hole states and $\nnbar$ states,
the only particle-hole states are observed at low excitation-energy
region. Thus, the relativistic corrections come from
the contribution of the $\nnbar$ states to Eq.(\ref{rel}).
They are estimated by calculating
the first term of Eq.(\ref{beta-})
for the total $\beta_-$ strength of the particle-hole states\cite{ksg1,ksg2},
\begin{eqnarray}
S_{{\rm ph}}=\frac{A}{\kf^3}\left\{Q(k_n)-Q(k_p)\right\},
\end{eqnarray}
where $q(k_i)$ is defined as
\begin{eqnarray}
Q(k_i)=\frac{k_i^3}{3}+2k_iM^{\ast 2}-2M^{\ast 3}\tan^{-1}\frac{k_i}{M^\ast}.
\end{eqnarray}
When we expand $Q(k_i)$ in terms of $(k_n-k_p)$, we obtain
\begin{eqnarray}
S_{{\rm ph}}\approx \left(1-\frac{2}{3}\vf^2\right)2(N-Z),
\end{eqnarray}
$\vf$ being the Fermi velocity, $\kf/\sqrt{M^{\ast 2}+\kf^2}$.
The reduction factor in the first parentheses is the relativistic correction
coming from the anti-nucleon degrees of freedom.
When we employ the values as $\vf=0.43$ which
corresponds to the values $M^\ast=0.6M$ and $\kf=1.36$ fm$^{-1}$ as in most
relativistic models\cite{ring},
the sum rule value is quenched by about 12\%.
The quenched amount is taken by $\nnbar$ states,
\begin{eqnarray}
[\mtrix{0}{F_+F_-}{0}-\mtrix{0}{F_-F_+}{0}]_{\nnbar}
\approx \frac{2}{3}\vf^22(N-Z).
\end{eqnarray}

In finite nuclei, we can not obtain the analytic formulae, but can
perform more sophisticated calculations numerically. We obtain
the quenching of the sum rule value by
6.3\% in $^{40}$Ca with use of the NL-SH parameter set\cite{ksg2},
and by 7.7\% in
$^{90}$Zr and by 8.4\% in $^{208}$Pb using the NL3\cite{ma}.
These values are smaller than 12\% in nuclear matter,
since the effective mass in nuclear
surface of finite nuclei is larger than that of nuclear matter.
These results implies that the relativistic
correction is not negligible in the discussion of the quenching
phenomena of the GT sum rule value.

\section{Relativistic RPA}

When we discuss the excitation strengths, the particle-hole correlations
should be examined as in non-relativistic models. Moreover,
the coupling between the particle-hole states and $\nnbar$ states
should be also investigated, since the present relativistic model
has no reason why the anti-nucleon degrees of freedom can be neglected.
The small coupling has a possibility to change the strengths
of low lying states,
because of the total strength of the $\nnbar$ states to be infinite.
To estimate these effects, we study the GT response function of nuclear
matter with the relativistic random phase
approximation(RPA)\cite{ks1,ks4}. This framework requires
two assumptions. First, we have to
assume the coupling Lagrangian between particle-hole states and
$\nnbar$ states. Second, we need a model to renormalize the
divergence in the response function. 

\subsection{RPA without renormalization}
For the coupling Lagrangian, we extend the non-relativistic
models for the giant GT states. In non-relativistic models, the
spin-isospin responses of nuclei are well described by the coupling
Hamiltonian\cite{brown},
\begin{eqnarray} 
V&=&\left[\left(\frac{f_\pi}{m_\pi}\right)^2g'\,
     \vct{\sigma}_1\cdot\vct{\sigma}_2\right.\nonumber\\
& &\nonumber\\
&-&\left.\left(\frac{f_\pi}{m_\pi}\right)^2 
 \frac{(\vct{\sigma}_1\cdot\vct{q})(\vct{\sigma}_2\cdot\vct{q})}{\vct{q}^2 
 +m_\pi^2} 
-\left(\frac{f_\rho}{m_\rho}\right)^2 
 \frac{(\vct{\sigma}_1\times\vct{q})(\vct{\sigma}_2\times\vct{q})}{\vct{q}^2 
 +m_\rho^2}\right]\vct{\tau}_1\cdot\vct{\tau}_2. \label{nonrelV}
\end{eqnarray}
The the first term with the Landau-Migdal parameter $g'$
is responsible for the short-range part of the interaction, the second
one is due to the pion exchange, and the last one comes from the
rho-meson exchange. In nuclear matter, only the first term is responsible
for the particle-hole interaction, since the momentum transfer is zero
in the GT excitations. In relativistic models,
we assume relativistic Landau-Migdal parameters which reduce to the
non-relativistic one. Those are given by the pseudovector and tensor
couplings\cite{mks},  
\begin{eqnarray}
 {\cal L}= 
 \frac{g_a}{2}\,
 \overline{\psi}\varGamma^\mu_i\psi\,
 \overline{\psi}\varGamma_{\mu i}\psi
 +
 \frac{g_t}{4}\,
 \overline{\psi}T^{\mu\nu}_i\psi\,
 \overline{\psi}T_{\mu\nu\, i}\,\psi,\label{relL}
\end{eqnarray}
where we have used the notations,
\begin{eqnarray}
\varGamma^\mu_i=\gamma_5\gamma^\mu\tau_i \,,\qquad
T^{\mu\nu}_i=\sigma^{\mu\nu}\tau_i. 
\end{eqnarray}
In the non-relativistic limit, both terms in Eq.(\ref{relL}) reduce to
the first term of Eq.(\ref{nonrelV}) with the relationship,
\begin{eqnarray}
g_a=g'\left(\frac{f_\pi}{m_\pi}\right)^2\,,\quad g_t
=g'\left(\frac{f_\pi}{m_\pi}\right)^2.
\end{eqnarray}
The role of the tensor coupling Lagrangian has been discussed in
ref.\cite{mks}, and shown to have no contribution to the quenching problems.
Therefore we neglect it from now on.

The relativistic RPA equation in nuclear matter
can be solved in an analytic way. When we
expand the excitation energy $\omega_{\rm GT}$
and strength $S_-$ of the GT state in terms of $(N-Z)$,
we have\cite{ks2} 
\begin{eqnarray}
\omega_{\rm GT} &\approx& \frac{4\kf^3}{3\pi^2}\,
 \frac{g_a}{1+2\kappa'(\nbar)g_a/(2\pi)^3}
 \left(1-\frac{2}{3}\vf^2\right)\, \frac{N-Z}{A}, \label{energy}\\
 & &\nonumber\\
S_- &\approx&
\frac{1}{\left( 1+2\kappa'(\nbar) g_a/(2\pi)^3 \right)^2}
\left(1-\frac23\vf^2\right)2\left(N-Z\right).\label{strength}
\end{eqnarray}
In the above equation,
$\kappa'(\nbar)$ stems from the coupling of the particle-hole
states with $\nnbar$ states,
\begin{eqnarray}
\kappa'(\nbar)&=&\frac23\int\!d^3p\,\frac{\vct{p}^2}{E_{\vct{p}}^3}
\left(2- \thn{p}-\thp{p} \right),\label{kappa}
\end{eqnarray}
which is infinite, because of the first term of the parentheses.
Thus, we need to renormalize the divergence in a proper way in
order to obtain the finite value. Before
discussing the renormalization, three comments will be in order.

First if we neglect the coupling between the particle-hole states and
$\nnbar$ states, the excitation strength of the GT state becomes
\begin{eqnarray}
S_- \longrightarrow
\left(1-\frac23\vf^2\right)2\left(N-Z\right),\label{strength2}
\end{eqnarray}
which exhausts the GT sum value of the nucleon sector. Moreover, 
in the non-relativistic limit, Eqs.(\ref{energy}) and (\ref{strength2})
reduce to 
\begin{eqnarray}
 \omega_{\rm GT}& \longrightarrow&  \frac{4\kf^3}{3\pi^2}
 g_a\,\frac{N-Z}{A}\,,\quad g_a=g'\left(\frac{f_\pi}{m_\pi}\right)^2,
\nonumber\\
S_-&\longrightarrow& 2(N-Z),\nonumber
\end{eqnarray}
which were obtained previously in non-relativistic models\cite{ts}.

Second,
as seen in Eqs.(\ref{energy}), the coupling constant $g_a$ is reduced by
the repulsive coupling with $\nnbar$ as
\begin{eqnarray}
g_a^{{\rm eff}}=\frac{g_a}{1+2\kappa'(\nbar)g_a/(2\pi)^3}.
\label{effective}
\end{eqnarray}
The strength is also reduced by the coupling, as seen in Eq.(\ref{strength}).
These forms are familiar in non-relativistic models where low lying states
couple with high lying ones by a repulsive force\cite{sk},
although the value of $\kappa'(\nbar)$ is positive infinite.

Third comment is rather serious for the approximation which neglects
simply the divergence terms, as in the the no-sea
approximation\cite{fruns}. This approximation has been extensively used
in previous calculations
to avoid the renormalization\cite{ksg1,ks2,ma,maring}.  
If the first term of Eq.(\ref{kappa}) is
neglected in order to avoid the divergence and the second and the
third term are kept, the positive $\kappa'(\nbar)$ is replaced
by the negative $\kappa(\nbar)$,
\begin{eqnarray}
\kappa(\nbar)&=&-\frac23\int\!d^3p\,\frac{\vct{p}^2}{E_{\vct{p}}^3}
\left(\thn{p}+\thp{p} \right).\label{kap}
\end{eqnarray}
Therefore, in this approximation, the repulsive
coupling is replaced artificially
by the attractive one. As a result, the excitation
energy becomes higher and the strength is enhanced due to the coupling
in spite of the fact that the coupling Hamiltonian is repulsive.

These unphysical results are
 also seen in the relativistic Landau-Migdal
parameter $F_0$ obtained previously from the same approximation in the
$\sigma$-$\omega$ model\cite{ks000,ks0}. 
When the coupling of the particle-hole states
with the $\nnbar$ states is neglected,
the $\sigma$-$\omega$ model provides us with $F_0$ as
\begin{eqnarray}
F_0=F_{\rm v} - (1-\vf^2)F_{\rm s},
  \label{nf0}
\end{eqnarray}
where we have defined
\begin{eqnarray}
 F_{\rm v}= N_{\rm F}\frac{g^2_{\rm v}}{m^2_{\rm v}},
\quad  F_{\rm s}=N_{\rm F}\frac{g^2_{\rm s}}{m^2_{\rm s}},
\quad N_{\rm F}=\frac{2\kf\ekf}{\pi^2}, \quad \ekf=\sqrt{\kf^2 + M^{*2}}.
\end{eqnarray}
In the above equations, $m_{\rm v(s)}$ and $g_{\rm v(s)}$
stand for the $\omega(\sigma)$-meson mass and
the coupling constant of the $\omega(\sigma)$-meson with the nucleon,
respectively. When we take into account the coupling with $\nnbar$
states by neglecting the divergence terms and keeping the Pauli blocking
terms, the Landau-Migdal parameter $F_0$ becomes
\begin{eqnarray}
F_0=F_{\rm v} - \frac{1-\vf^2}{1+a_{\rm s}F_{\rm s}}F_{\rm s},
\label{f0}
\end{eqnarray}
where we have used the abbreviations,
\begin{eqnarray}
a_{\rm s} = \frac{3}{2}
 \left( 1- \frac{2}{3}v_{\rm F}^2 + \frac{1-v_{\rm F}^2}{2v_{\rm F}}
  \ln \frac{1-v_{\rm F}}{1+v_{\rm F}}\right)
\approx \frac{1}{5}\vf^4\left(1+\frac{3}{7}\vf^2+\cdots\right).
\end{eqnarray}
Thus, the attractive part due to the $\sigma$-meson exchange is
un-physically reduced due to the coupling with the $\nnbar$ states by
the factor $(1+a_{\rm s}F_{\rm s})>0$, in spite of the fact that
the attractive part should be enhanced by the coupling.
The Landau-Migdal parameter $F_0$
dominates the excitation energy of the giant monopole states. The
reduction of the attractive force yields the increase of the excitation
energy\cite{ks0}. Recent numerical calculations of the excitation energy
have also shown this fact,
although the experimental values are well reproduced\cite{maring}. 

It is worthwhile noting one more comment on the approximation
to neglect the divergence terms. Even if one neglects the divergence
terms, the GT sum rule is satisfied,
since the infinite terms in Eqs.(\ref{beta-}) and (\ref{beta+}) exactly 
cancel each other. The analytic formulae in Eqs. (\ref{beta-})
and (\ref{beta+}), however,
show that in this approximation, the $\beta_\pm$ transition
strengths of $\nnbar$ states
become negative. This fact changes the effect
of the repulsive interaction into that of
attractive one un-physically. We have also the negative strengths above 1GeV
in the RPA calculations\cite{ks00}.
Thus the sum rule is not enough for justification of the approximation.

\subsection{RPA with renormalization}
Now let us discuss the renormalization of the divergence on the first
term of Eq.(\ref{kappa}). For this purpose, we employ $n$-dimensional
regularization method\cite{chin}. 
The 3-dimensional integral of the first term came from the
4-dimensional one,
\begin{eqnarray}
\frac23\int\!d^3p\,\frac{\vct{p}^2}{E_{\vct{p}}^3}2
&=&i\frac{2}{\pi}\int d^4p\frac{M^\ast+p^2+2p_y^2}{(p^2-M^{\ast 2}+i\epsilon)^2}.
\end{eqnarray}
The above 4-dimensional integral is extended to the
$n$-dimensional integral,
\begin{eqnarray}
I_n=i\frac{2}{\pi}\int d^np\frac{M^\ast+p^2+2p_y^2}{(p^2-M^{\ast 2}+i\epsilon)^2}.
\end{eqnarray}
In the limit $n\rightarrow 4$,
we can separate the integral into the
finite part and infinite part,
\begin{eqnarray}   
I_{n\rightarrow 4}=4\pi\left\{M^{\ast 2}\Gamma (1-n/2)|_{n\rightarrow 4}+2M^{\ast 2}\ln M^\ast\right\},
\end{eqnarray}
where the divergence is included in the $\Gamma$ function.
Since the $\Gamma$ function is multiplied by $M^{\ast 2}$,
\begin{eqnarray}
M^{\ast 2}=(M-U_s)^2=M^2-2MU_s+U^2_s,\nonumber
\end{eqnarray}
we assume the 3 counter terms to cancel the divergence
in the Lagrangian\cite{mks2},
\begin{eqnarray}
{\Delta {\cal L}_c
 = \frac{1}{2}\left\{a_0+a_1\sigma
  +\frac{1}{2}a_2\sigma^2 \right\} \vct{A}_\mu\!\cdot\!\vct{A}^\mu}.
\end{eqnarray}
In the above equation, $\sigma$ and $\vct{A}_\mu$ denote the Lorentz scalar
and pseudovector fields, respectively, and the coefficients $a_i$ are
determined so as to cancel the divergence.
Finally, we obtain the renormalized $\kappa'(\nbar)$ as
\begin{eqnarray}
\kappa'(\nbar)_{{\rm ren}}&=&
-4\pi M^2\left\{2 \frac{M^{\ast\,2}}{M^2} \ln \frac{M}{M^\ast}
+\left(1-\frac{M^\ast}{M}\right)\left(1-3\frac{M^\ast}{M}\right)
\right\}
 -\frac{16\pi}{15}\kf^2\vf^3 \nonumber\\
&\approx& -4\pi M^2 \frac23 \left( 1-\frac{M^\ast}{M} \right)^3
 -\frac{16\pi}{15}\kf^2\vf^3.
 \label{ren-kappa}
\end{eqnarray}
The first term of the right hand side is obtained by the renormalization
of the divergence, and the last term comes from the Pauli blocking one.
It should be noted that the above $\kappa'(\nbar)_{{\rm ren}}$ is
negative for $M^\ast<M$, while $\kappa'(\nbar)$ in Eq.(\ref{kappa}) is
positive infinite. Thus, the renormalization by the counter terms
change the repulsive contribution 
from $\nnbar$ states into the attractive one to Eq.(\ref{effective}).
This change happens in other cases, for example, in the renormalization
of the divergence in the Landau-Migdal parameter $F_0$\cite{ks0}. 

Let us estimate the quenching of the GT strength
by using $g'=0.6$, $M^\ast= 0.7306M$ and $\kf=1.3$ fm$^{-1}$.
The first value is determined so as to reproduce the excitation energy of
the GT state\cite{ss}, while the last two ones are obtained
in the renormalized Hartree approximation of the $\sigma-\omega$
model to explain the binding energy of nuclear matter\cite{ks4}.  
Finally we obtain 
\begin{eqnarray}
S_-=\frac{1-2\vf^2/3}
{\left\{1+2\kappa'(\nbar)_{{\rm ren}}g_a/(2\pi)^3\right\}^2}
2(N-Z)\approx 2(N-Z).
\end{eqnarray}
Thus, the renormalization is very important, and almost cancels the
quenching in the Hartree approximation.
For this result, the first term in Eq.(\ref{ren-kappa}) is
important, and second term from the Pauli blocking one is negligible.
In fact, if we neglect simply the divergence terms in the RPA, we have
\begin{eqnarray}
S_-=\frac{1-2\vf^2/3}
 {\left\{1+2\kappa(\nbar)g_a/(2\pi)^3\right\}^2}
2(N-Z)\approx \left(1-\frac23\vf^2\right)2(N-Z),
\end{eqnarray}
as shown in refs.\cite{ksg1} and \cite{ksg2}.

We note that $\kappa'(\nbar)_{{\rm ren}}$ has a rather strong
density-dependence through the effective mass. In finite nuclei,
the effect of $\kappa'(\nbar)_{{\rm ren}}$ on $S_-$ may be weaken.
In high density nuclear matter, on the contrary, the value of $S_-$ is
expected to be more enhanced.

Before closing this subsection, two comments should be added.
The relativistic RPA satisfies the GT sum rule. Second, the
energy-weighted sum of the $\beta_-$ and $\beta_+$ transition strengths
is equal to the ground-state expectation value of the double
commutator as to the GT operator and  the present
Hartree Hamiltonian\cite{ks2},
\begin{eqnarray}
& &\sum_n(E_n-E_0)|\mtrix{n}{F_-}{\tilde{0}}|^2
 +\sum_n(E_n-E_0)|\mtrix{n}{F_+}{\tilde{0}}|^2 \nonumber \\
&=&
\mtrix{0}{[F_+,\,[H, F_-]]}{0},
\end{eqnarray}
where $E_n$ and $E_0$ stand for the excitation energies of the RPA excited
state $\ket{n}$ and the ground state $\ket{\tilde{0}}$, respectively,
and $H$ denotes
the Hartree Hamiltonian. The above relationship is well known in
non-relativistic RPA for no-charge exchange excitations\cite{thouless}.

\section{Conclusion}

The Gamow-Teller(GT) sum rules with respect to the
difference between $\beta_-$ and $\beta_+$ transition strengths
are investigated in the relativistic and non-relativistic theories.
The sum rule value of the relativistic theory is the same as that of the
non-relativistic theory, although each total strength of
the $\beta_-$ and $\beta_+$ transitions is infinite.
In the non-relativistic theory, the sum rule value is exhausted
by the particle-hole states, while in the relativistic theory,
a part of the GT strengths are taken
by the nucleon-antinucleon($\nnbar$) states.
Hence, if there is no coupling between the particle-hole states
and $\nnbar$ states, the total GT strengths of the particle-hole
states in the relativistic theory are reduced,
compared to the sum rule value.
According to the relativistic nuclear models developed
for recent years\cite{serot,ring},
the quenching amount is estimated to be about 12\% of the sum rule value
in nuclear matter, and about 8\% in finite nuclei

There may be a possibility that the coupling of the particle-hole states
can be neglected, from more fundamental reason beyond the present
relativistic model. We have estimated, however, effects of the coupling
on the GT strengths within the relativistic model, using the
random phase approximation for nuclear matter. The divergence of the GT
response function due to the $\nnbar$ states is properly
renormalized by the $n$-dimensional regularization method\cite{chin}.
It has been shown that the coupling reduces the quenching
of the GT strengths. In nuclear matter with the normal density,
it is expected that the total strength of the
$\beta_-$ transition to the low lying states becomes nearly equal
to the GT sum rule value. Since the coupling effect is density-dependent,
it may be weaken in finite nuclei. In this sense, we expect that 
8\%  is the upper limit of the relativistic correction in the present
relativistic model.

We have also discussed the approximation like the no-sea approximation
which neglects the divergent terms,
but takes into account the Pauli blocking ones of the $\nnbar$
excitations. The approximation does not violate the GT sum rule, but
yields negative excitation strengths for $\nnbar$ states. Because of
this fact, the repulsive interaction of the coupling works,
as if it were an attractive one.
We have the same problem in a description of the giant monopole
states, where the attractive coupling interaction through the $\sigma$-meson
exchange yields a repulsive effect. There seems to be no justification
for the approximation to neglect the divergence.     

\section*{Acknowledgments}
The authors would like to thank Drs. N. Van Giai, Z.-Y. Ma and
T. Maruyama for useful discussions.


\begin{thebibliography}{99}
\bibitem{yako}K. Yako et al., Phys. Lett. B {\bf 615}, 193 (2005).
\bibitem{sts}T. Suzuki, H. Sakai, and T. Tatsumi, in proceedings of the
	RCNP International Symposium on Nuclear Response and Medium
	Effects, edited by T. Noro et al., p.77 (Universal Academic
	Press, Tokyo, 1999).
\bibitem{ss}T. Suzuki and H. Sakai, Phys. Lett. B {\bf 455}, 25 (1999). 
\bibitem{bentz}A. Arima, et al., Phys. Lett. B {\bf 499}, 104 (2001).
\bibitem{ksg1}H. Kurasawa, T. Suzuki and N. Van Giai,
        Phys. Rev. Lett. {\bf 91}, 062501 (2003).
\bibitem{ksg2}H. Kurasawa, T. Suzuki and N. Van Giai, Phys. Rev. {\bf
	C68}, 064311 (2003).
\bibitem{ksg3}H. Kurasawa, T. Suzuki and N. Van Giai, Nucl. Phys.
	{\bf A731}, 114 (2004)
\bibitem{ks2}H. Kurasawa and T. Suzuki,
	Phys. Rev. C {\bf 69}, 014306 (2004).
\bibitem{ma}Z.-Y. Ma, et al., Euro. J. Phys. {\bf A20}, 429 (2004).
\bibitem{ks00}H. Kurasawa and T. Suzuki, Phys. of Atomic Nucl. {\bf 67},
	1656 (2004).
\bibitem{mks}T. Maruyama, H. Kurasawa and T. Suzuki,
	Prog. Theor. Phys. {\bf 113}, 355 (2005).
\bibitem{fruns}J. F. Dawson and R. J. Furnstahl, Phys. Rev. {\bf C42},
	2009 (1990).
\bibitem{maring}Z.-Y. Ma et al., Nucl. Phys. {\bf A722}, C491 (2003),
	and references therein.
\bibitem{osterfeld}F. Osterfeld, Rev. Mod. Phys. {\bf 64}, 491 (1992).
\bibitem{serot}B. D. Serot and J. D. Walecka, Adv. Nucl. Phys.
	{\bf 16}, 1 (1986).
\bibitem{ring}P. Ring, Prog. Part. Nucl. Phys. {\bf 37}, 193 (1996).
\bibitem{ks1}H. Kurasawa and T. Suzuki, Nucl. Phys. {\bf A445}, 685 (1985).
\bibitem{ks4}H. Kurasawa and T. Suzuki, Nucl. Phys. {\bf A490}, 571 (1988).
\bibitem{brown} S. -O. B$\ddot{{\rm a}}$ckman, G. E. Brown,
	and J. A. Niskanen,
	Phys. Rep. {\bf 24}, 1 (1985).
\bibitem{ts}T. Suzuki, Nucl. Phys. {\bf A379}, 110 (1982).
\bibitem{sk}T. Suzuki and M. Kohno, Prog. Theor. Phys. {\bf 68}, 690 (1982).
\bibitem{ks000}H. Kurasawa and T. Suzuki, Nucl. Phys. {\bf A454}, 527 (1986).
\bibitem{ks0}H. Kurasawa and T. Suzuki, Phys. Lett. {\bf B474}, 262 (2000).
\bibitem{chin}S. A. Chin, Ann. Phys. (N.Y.) {\bf 108}, 301 (1977).
\bibitem{mks2}T. Maruyama, H. Kurasawa and T. Suzuki, to be published.
\bibitem{thouless}D. J. Thouless, Nucl. Phys. {\bf 22}, 78 (1961).
\end{thebibliography}
\end{document}